\documentclass[10pt,conference,english]{IEEEtran}
\IEEEoverridecommandlockouts
\usepackage[centertags]{amsmath}
\usepackage{amsfonts}
\usepackage{amssymb}
\usepackage{epsfig}
\usepackage{amsmath,  amssymb}
\usepackage{graphicx}
\usepackage[centertags]{amsmath}
\usepackage{clrscode}
\usepackage{cite}
\usepackage{fullpage}
\usepackage{amsmath,amssymb,mathrsfs,pseudocode}
\usepackage{color}
\usepackage[normalem]{ulem}
\usepackage{psfrag}
\usepackage{url}
\usepackage{babel}
\usepackage{authblk}
\usepackage{mathtools}
\usepackage[linesnumbered,ruled]{algorithm2e}
\usepackage[left=54pt, right=54pt, top=51pt, bottom=58pt]{geometry}
\usepackage{xspace}
\usepackage{multicol}
\usepackage{booktabs} 
\usepackage{multirow}
\usepackage[x11names,dvipsnames,table]{xcolor}
\usepackage{setspace}

\usepackage{subfigure}

\usepackage{stackrel}
\usepackage{extarrows}
\usepackage{bbm}

\newtheorem{theorem}{Theorem}
\newtheorem{lemma}{Lemma}

\newtheorem{proposition}{Proposition}
\newtheorem{example}{Example}
\newtheorem{definition}{Definition}
\newtheorem{corollary}{Corollary}
\newtheorem{remark}{Remark}

\usepackage{xspace}
\usepackage{bbm}
\catcode`@=11
\chardef\mathlig@atcode\count255

\def\actively#1#2{\begingroup\uccode`\~=`#2\relax\uppercase{\endgroup#1~}}
\def\mathlig@gobble{\afterassignment\mathlig@next@cmd\let\mathlig@next= }
\def\mathlig@delim{\mathlig@delim}
\def\mathlig@defcs#1{\expandafter\def\csname#1\endcsname}
\def\mathlig@let@cs#1#2{\expandafter\let\expandafter#1\csname#2\endcsname}
\def\mathlig@appendcs#1#2{\expandafter\edef\csname#1\endcsname{\csname#1\endcsname#2}}

\def\mathlig#1#2{\mathlig@checklig#1\mathlig@end\mathlig@defcs{mathlig@back@#1}{#2}\ignorespaces}

\def\mathlig@checklig#1#2\mathlig@end{%
 \expandafter\ifx\csname mathlig@forw@#1\endcsname\relax
 \expandafter\mathchardef\csname mathlig@back@#1\endcsname=\mathcode`#1%
 \mathcode`#1"8000\actively\def#1{\csname mathlig@look@#1\endcsname}%
 \mathlig@dolig#1\mathlig@delim
\fi
\mathlig@checksuffix#1#2\mathlig@end
}

\def\mathlig@checksuffix#1#2\mathlig@end{%
\ifx\mathlig@delim#2\mathlig@delim\relax\else\mathlig@checksuffix@{#1}#2\mathlig@end\fi
}
\def\mathlig@checksuffix@#1#2#3\mathlig@end{%
\expandafter\ifx\csname mathlig@forw@#1#2\endcsname\relax\mathlig@dosuffix{#1}{#2}\fi
\mathlig@checksuffix{#1#2}#3\mathlig@end
}

\def\mathlig@dosuffix#1#2{%
\mathlig@appendcs{mathlig@toks@#1}{#2}%
\mathlig@dolig{#1}{#2}\mathlig@delim
}

\def\mathlig@dolig#1#2\mathlig@delim{%
 \mathlig@defcs{mathlig@look@#1#2}{%
 \mathlig@let@cs\mathlig@next{mathlig@forw@#1#2}\futurelet\mathlig@next@tok\mathlig@next}%
 \mathlig@defcs{mathlig@forw@#1#2}{%
  \mathlig@let@cs\mathlig@next{mathlig@back@#1#2}%
  \mathlig@let@cs\checker{mathlig@chck@#1#2}%
  \mathlig@let@cs\mathligtoks{mathlig@toks@#1#2}%
  \expandafter\ifx\expandafter\mathlig@delim\mathligtoks\mathlig@delim\relax\else
  \expandafter\checker\mathligtoks\mathlig@delim\fi
  \mathlig@next
 }%
 \mathlig@defcs{mathlig@toks@#1#2}{}%
 \mathlig@defcs{mathlig@chck@#1#2}##1##2\mathlig@delim{%
  \ifx\mathlig@next@tok##1%
   \mathlig@let@cs\mathlig@next@cmd{mathlig@look@#1#2##1}\let\mathlig@next\mathlig@gobble
  \fi 
  \ifx\mathlig@delim##2\mathlig@delim\relax\else
   \csname mathlig@chck@#1#2\endcsname##2\mathlig@delim
  \fi
 }%
 \ifx\mathlig@delim#2\mathlig@delim\else
  \mathlig@defcs{mathlig@back@#1#2}{\csname mathlig@back@#1\endcsname #2}%
 \fi
}%

\catcode`@\mathlig@atcode

\newcommand{\muspace}{\mspace{1mu}}

\DeclareRobustCommand{\scond}{\mathchoice{\muspace\vert\muspace}{\vert}{\vert}{\vert}}
\mathlig{|}{\scond}

\DeclareRobustCommand{\discint}{\mathchoice{\mspace{-1.5mu}:\mspace{-1.5mu}}{\mspace{-1.5mu}:\mspace{-1.5mu}}{:}{:}}
\mathlig{::}{\discint}

\newcommand{\Ec}{\mathcal{E}}

\newcommand{\Hc}{\mathcal{H}}
\newcommand{\Ic}{\mathcal{I}}

\newcommand{\Kc}{\mathcal{K}}
\newcommand{\Lc}{\mathcal{L}}
\newcommand{\Mc}{\mathcal{M}}

\newcommand{\Rc}{\mathcal{R}}
\newcommand{\Sc}{\mathcal{S}}
\newcommand{\Tc}{\mathcal{T}}

\newcommand{\Xc}{\mathcal{X}}
\newcommand{\Yc}{\mathcal{Y}}
\newcommand{\Zc}{\mathcal{Z}}

\newcommand{\Gcal}{\mathcal{G}}

\newcommand{\Kcal}{\mathcal{K}}

\newcommand{\Pcal}{\mathcal{P}}

\newcommand{\Xcal}{\mathcal{X}}

\newcommand{\Zcal}{\mathcal{Z}}

\newcommand{\Av}{{\bf A}}
\newcommand{\Bv}{{\bf B}}

\newcommand{\Dv}{{\bf D}}
\newcommand{\Ev}{{\bf E}}

\newcommand{\Kv}{{\bf K}}

\newcommand{\Mv}{{\bf M}}
\newcommand{\Xv}{{\bf X}}
\newcommand{\Evt}{{\tilde{\bf E}}}
\newcommand{\Kvt}{{\tilde{\bf K}}}
\newcommand{\Mvt}{{\tilde{\bf M}}}

\newcommand{\Svt}{{\tilde{\bf S}}}
\newcommand{\Uvt}{{\tilde{\bf U}}}

\newcommand{\Yv}{{\bf Y}}

\newcommand{\Uv}{{\bf U}}
\newcommand{\Wv}{{\bf W}}

\newcommand{\Sv}{{\bf S}}

\newcommand{\dv}{{\bf d}}

\newcommand{\gv}{{\bf g}}

\newcommand{\xv}{{x}}

\newcommand{\Qh}{{\hat{Q}}}

\newcommand{\Yh}{{\hat{Y}}}

\newcommand{\Qt}{{\tilde{Q}}}

\newcommand{\Yt}{{\tilde{Y}}}

\newcommand{\xt}{{\tilde{x}}}

\def\d{\delta}
\def\e{\epsilon}

\def\textiid{i.i.d.\@\xspace}
\newcommand\iid{\ifmmode\text{ i.i.d. } \else \textiid \fi}

\def\clap#1{\hbox to 0pt{\hss#1\hss}}
\def\mathclap{\mathpalette\mathclapinternal}
\def\mathclapinternal#1#2{%
  \clap{$\mathsurround=0pt#1{#2}$}}

\let\oldstackrel\stackrel
\renewcommand{\stackrel}[2]{\oldstackrel{\mathclap{#1}}{#2}}

\newcommand{\parastoo}{\textcolor{black}}

\newcommand{\Roy}{\textcolor{red}}

\newcommand{\Royy}{\textcolor{blue}}

\newcommand{\Roycr}{\textcolor{black}}

\newcommand{\Rarxiv}{\textcolor{black}}

\newcommand{\leak}{L}
\newcommand{\Fb}{{\mathbb F}}

\allowdisplaybreaks
\begin{document}

\title{Information Leakage in Index Coding With Sensitive and Non-Sensitive Messages}


\author{Yucheng Liu$^{\dag}$\thanks{This work was supported by the ARC Discovery Scheme DP190100770; the US National Science Foundation Grant CNS-1815322; and the ARC Future Fellowship FT190100429.}, Lawrence Ong$^{\dag}$, Phee Lep Yeoh$^{\S}$, Parastoo Sadeghi$^{\ddag}$, Joerg Kliewer$^{*}$, and Sarah Johnson$^{\dag}$, \\\vspace{-0mm}
$^{\dag}$The University of Newcastle, Australia (emails: \{yucheng.liu, \hspace{-0.5mm}lawrence.ong, \hspace{-0.5mm}sarah.johnson\}@newcastle.edu.au)\\
$^{\S}$University of Sydney, Australia (email: phee.yeoh@sydney.edu.au)\\
$^{\ddag}$University of New South Wales, Canberra, Australia (email: p.sadeghi@unsw.edu.au)\\
$^{*}$New Jersey Institute of Technology, USA (email: jkliewer@njit.edu)\\\vspace{0mm}
}




\maketitle


\begin{abstract}

Information leakage to a guessing adversary in index coding is studied, where some messages in the system are sensitive and others are not. 
The non-sensitive messages can be used by the server like secret keys to mitigate leakage of the sensitive messages to the adversary. 
We construct a deterministic linear coding scheme, developed from the rank minimization method based on fitting matrices (Bar-Yossef et al. 2011). 
The linear scheme leads to a novel upper bound on the optimal information leakage rate, which is proved to be tight over all deterministic scalar linear codes. 
We also derive a converse result from a graph-theoretic perspective, which holds in general over all deterministic and stochastic coding schemes.

\end{abstract}

\section{Introduction}\label{sec:intro}

Index coding \cite{Birk--Kol1998,bar2011index} is a canonical problem in network information theory, where a server tries to efficiently broadcast messages to multiple receivers with side information. 
In this work, we study the information leakage in index coding when the broadcast codeword is eavesdropped by a guessing adversary. 
The adversary tries to maximize the probability of correctly guessing its messages of interest within a single trial. 
Our goal is to minimize the leakage to this adversary, which is defined as the ratio between the adversary's probability of successful guessing \emph{after and before} observing the codeword \cite{smith2009foundations,braun2009quantitative,issa2019operational}.

\begin{figure}[ht]
\begin{center}
\includegraphics[scale=0.175]{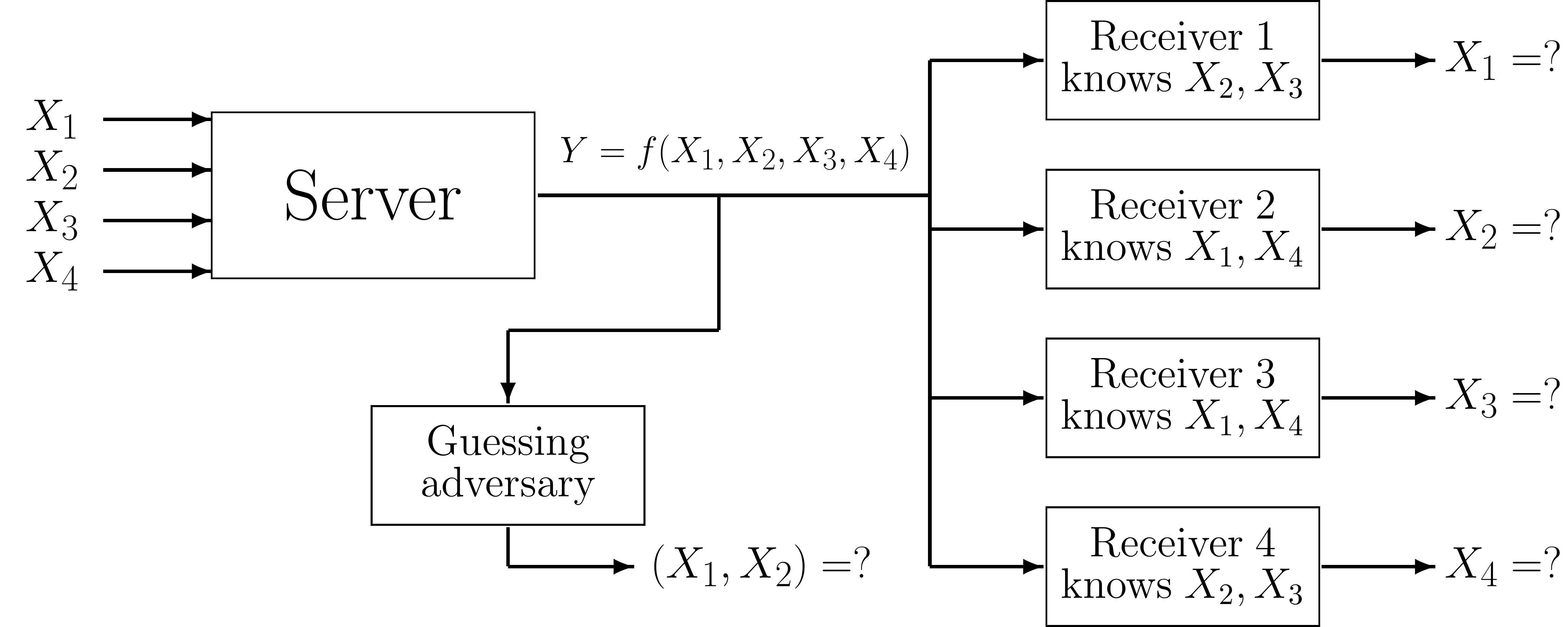}
\caption{There are four binary messages $X_i,i\in [4]$, where $X_1,X_2$ are sensitive and $X_3,X_4$ are non-sensitive. 
A guessing adversary eavesdrops the broadcast codeword $Y$ and tries to guess the sensitive messages. 
When guessing blindly (without knowing $Y$), the probability of the adversary correctly guessing $(X_1,X_2)$ is only $1/4$. 
To satisfy the legitimate receivers, the server can generate $Y$ as $Y=(X_1\oplus X_2,X_3\oplus X_4)$. 
However, such $Y$ leads to certain amount of information leakage as the adversary's correct guessing probability when observing it becomes $1/2$. 
To simultaneously satisfy the receivers and prevent any information leakage, the server can broadcast $\Yt=(X_1\oplus X_3,X_2\oplus X_4)$. 
In this way, the sensitive messages are protected against the adversary using non-sensitive messages. 
\vspace{0mm}}
\label{fig:index:coding:with:adversary}
\end{center}
\end{figure}

Our previous work \cite{itw2021leakage} studied the information leakage with an underlying assumption that the server aims to protect \emph{all} of the messages against the adversary. 
Indeed, such assumption holds in most existing works \cite{dau2012security,ong2016secure,ong2018secure,mojahedian2017perfectly,liu:vellambi:kim:sadeghi:itw18} where the security and privacy aspects of index coding are investigated. 
Nevertheless, in many practical circumstances, some messages may be sensitive while the others are non-sensitive, and thus secrecy loss should be measured over only those sensitive messages. 
For example, consider a data storage system with a number of messages. Some messages are private and needs to be protected from any adversary, while the other messages are non-private and thus need not be protected. There are a number of clients, each of which may request access to some messages, whether private or non-private, and may have stored some other messages already as side information. The goal is to satisfy the clients’ requirements while keeping the private messages as safe as possible from the adversary. 
A similar setting can also be motivated from the adversary's perspective as it may only be interested in a subset of messages and thus these messages should be considered as sensitive by the server. 
This distinction between sensitive and non-sensitive messages enables the server to treat the non-sensitive messages like secret keys and design a smart way of coding to simultaneously satisfy the receivers and mitigate information leakage to the adversary. 
%
Figure \ref{fig:index:coding:with:adversary} serves as a toy example showing how the server can reduce the information leakage by smartly designing coding schemes.  



\vspace{-2pt}

We study the problem of both achievability and converse. 
In Section \ref{sec:model}, we describe the system model and provide necessary preliminaries. 
In Section \ref{sec:achievability}, we propose a practical linear coding scheme based on the rank minimization method over fitting matrices \cite{bar2011index}. The scheme is proved to yield optimal (minimum) information leakage rate over all deterministic scalar linear codes. 
In Section \ref{sec:converse}, we develop a general lower bound on the optimal leakage rate that holds over any  coding schemes, even stochastic ones. 
To show the lower bound, we use the confusion graph representation of index coding problems \cite{alon2008broadcasting} and apply graph-theoretic tools. 

\section{Problem Formulation}\label{sec:model}

\subsection{System Model and Preliminaries}






We consider a server containing $n$ uniformly distributed and independent messages $X_i,i\in [n]$.
Each message is a sequence of length $t$ that takes values from $\Xc^t$ for some finite field $\Xc={\mathbb F}_q$. 
For any $S\subseteq [n]$, set $X_S\doteq (X_i,i\in S)$, $x_S\doteq (x_i,i\in S)$, and $\Xc_S\doteq \Xc^{|S|t}$. 
Thus $X_{[n]}$ denotes the tuple of all $n$ messages, and $x_{[n]}\in \Xc_{[n]}$ denotes a realization of the message $n$-tuple. 
By convention, $X_{\emptyset} = x_{\emptyset} = \Xc_{\emptyset} = \emptyset$. 

The server encodes the $n$ messages to some codeword $Y$ and broadcasts it to $m$ receivers via a noiseless channel. 
Receiver $i\in [m]$ wants to know messages $X_{W_i}$ and has $X_{A_i}$ as side information. 
We allow degenerated receivers in the system, who wants nothing (i.e., $W_i=\emptyset$). Such a receiver can always decode what it wants perfectly by definition. 

More formally, a $(t,M,f,\gv)$ {\em index code} is defined by
\begin{itemize}
\item One \emph{stochastic} encoder $f: \Xc^{nt} \to \{1,2,\ldots,M\}$ at the server that maps each message tuple $\xv_{[n]}\in \Xc^{nt}$ to a codeword $y\in \{1,2,\ldots,M\}$, and
\item $m$ \emph{deterministic} decoders $\gv=(g_i,i\in [m])$, one for each receiver $i \in [m]$, such that $g_i: \{1,2,\ldots,M\} \times \Xc^{|A_i|t} \to \Xc^{|W_i|t}$ maps the codeword $y$ and the side information $x_{A_i}$ to some estimated sequence $\hat{x}_{W_i}$.
\end{itemize}

We say a $(t,M,f,\gv)$ index code is \emph{valid} if and only if (iff) every receiver can perfectly decode its wanted messages. 
%
%
A compression rate $R$ is achievable iff there exists a valid $(t,M,f,\gv)$ code such that $R\ge (\log_q M)/t$. 
The {optimal} rate $\beta$, also called the \emph{broadcast rate}, can be defined as \cite{arbabjolfaei2018fundamentals}
\begin{align}  \label{eq:model:broadcast:rate}
\beta&=\lim_{t\to \infty}\min_{\substack{\text{valid $(t,M,f,\gv)$ code}}}\frac{\log_q M}{t}. 
\end{align}


%

\begin{remark}
Broadcast rate can also be defined allowing vanishing decoding error. 
However, for index coding, the zero-error and vanishing-error broadcast rates are equal \cite{Langberg--Effros2011}. 
\end{remark}


Any index coding instance is described by the parameter tuple $(n,m,(W_i,i\in [m]),(A_i,i\in [m]))$.  
 

\subsection{Confusion Graph}

Any index coding instance can also be characterized by 
a family of {confusion graphs}, $(\Gamma_t,t\in {\mathbb Z}^+)$ \cite{alon2008broadcasting}. 
For a given sequence length $t$, the confusion graph $\Gamma_t$ is an undirected graph defined on the message tuple alphabet $\Xc_{[n]}$. That is, the vertex set $V(\Gamma_t)=\Xc_{[n]}$. 
Vertex $x_{[n]}$ in $\Gamma_t$ corresponds to the realization $x_{[n]}$. 
Any two different vertices $x_{[n]},z_{[n]}$ are adjacent in $\Gamma_t$ iff there exists some receiver $i\in [m]$ such that $x_{W_i}\neq z_{W_i}$ and $x_{A_i}=z_{A_i}$. 
We call any pair of vertices satisfying this condition \emph{confusable} at receiver $i$, or simply confusable. 
Hence, the edge set $E(\Gamma_t)=\{ \{x^t_{[n]},z^t_{[n]}\}:\text{$x_{W_i}\neq z_{W_i}$ and $x_{A_i}=z_{A_i}$ for some $i\in [m]$} \}$. 

\begin{figure}[ht]
\begin{center}
\includegraphics[scale=0.35]{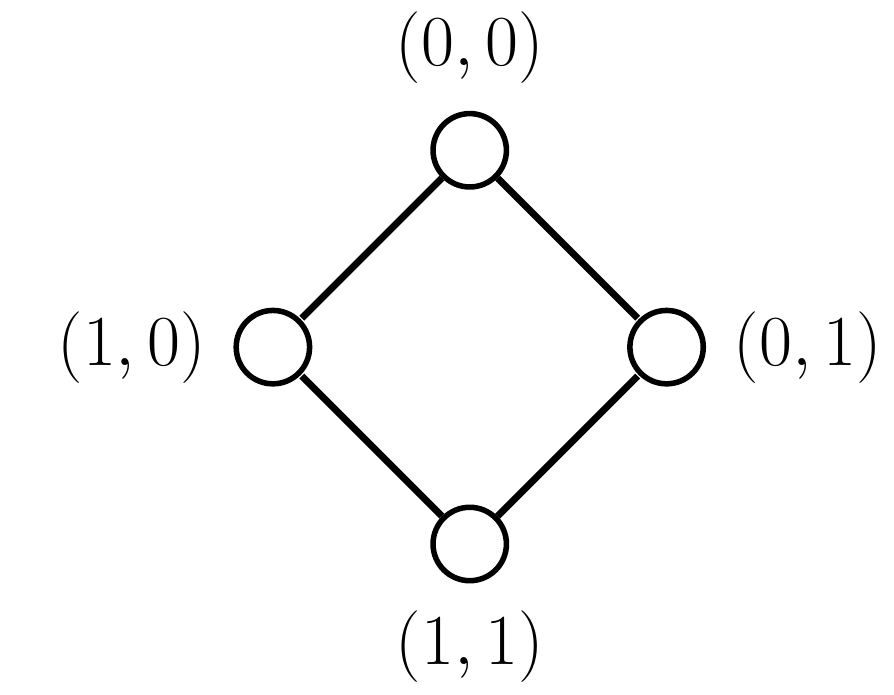}
\caption{The confusion graph $\Gamma_1$ with $t=1$ for the index coding instance $(2,2,(\{1\},\{2\}),(\{2\},\{1\}))$. 
Note that, for example, $x_{[n]}=(0,0)$ and $z_{[n]}=(0,1)$ are confusable at receiver $2$, because $x_{W_2}=x_{2}=0\neq z_{W_2}=z_{2}=1$ and $x_{A_2}=x_1=0=z_1=z_{A_2}$. 
Suppose $(0,0)$ and $(0,1)$ are mapped to the same codeword $y$ with certain nonzero probabilities. Then upon receiving this $y$, receiver $2$ will not be able to tell whether the value for $X_2$ is $0$ or $1$ based on its side information of $X_1=0$. 
}
\label{fig:confusion:graph}
\end{center}
\end{figure}

For correct decoding at all receivers, any two values $x_{[n]},z_{[n]}$ can be mapped to the same codeword $y$ with nonzero probabilities iff they are not confusable \cite{alon2008broadcasting}. See Figure \ref{fig:confusion:graph} below for a toy example of an index coding instance and its confusion graph. 
For the definitions for basic graph-theoretic notions, see any textbook on graph theory (e.g., Scheinerman and Ullman \cite{scheinerman2011fractional}).

\vspace{-2pt}

To simplify the notation, we may use $\Gamma$ to denote an index coding instance characterized by $(\Gamma_t,t\in {\mathbb Z}^+)$. 
We denote the broadcast rate of the problem $\Gamma$ as $\beta(\Gamma)$ when this dependence needs to be emphasized.

Consider any set $J\subseteq [n]$. 
The subproblem induced by message subset $J$ is characterized by the tuple $(|J|,m,(W_i\cap J,i\in [m]),(A_i\cap J,i\in [m]))$. 
Let $\Gamma(J)$ and $\Gamma_t(J)$ denote the subproblem induced by $J$ itself and the confusion graph of message length $t$ of the subproblem, respectively. 
The broadcast rate $\beta(\Gamma)$ can be characterized by the confusion graphs $(\Gamma_t,t\in {\mathbb Z}^+)$ as
\begin{align}
\beta(\Gamma)=\lim_{t\to \infty}\frac{1}{t}\log_q \chi(\Gamma_t)=\lim_{t\to \infty}\frac{1}{t}\log_q \chi_{\rm f}(\Gamma_t),  \label{eq:compression:zero:error}
\end{align}
where $\chi(\cdot)$ and $\chi_{\rm f}(\cdot)$ respectively denote the chromatic number and fractional chromatic number of a graph. 
The proof of \eqref{eq:compression:zero:error} can be found in \cite[Section 3.2]{arbabjolfaei2018fundamentals}. 

\subsection{Information Leakage Metric}

We assume the broadcast codeword is eavesdropped by a guessing adversary, knowing a subset of messages $X_K$ as side information. 
The rest of the messages $X_{K^c}$ are divided into two groups, where the sensitive messages are denoted by $X_S$ and the non-sensitive ones are denoted by $X_U$. 
The information leakage from $Y$ to the adversary will be measured only over the sensitive messages $X_S$. 
Upon observing the broadcast codeword, the adversary makes a single guess on the value of $X_S$ according to the maximum likelihood rule. 
Note that $K,S,U$ are non-overlapping and $[n]=K\cup S\cup U$. Let $k,s,u$ denote the cardinality of sets $K,S,U$, respectively.

Consider any valid $(t,M,f,\gv)$ index code. 
Before eavesdropping the codeword $Y$, the expected probability of the adversary successfully guessing $x_S$ is 
\begin{align*}
{\rm P_{\rm s}}(X_K)={\mathbb E}_{X_K} \left[ \max_{x_S} P_{X_S|X_K}(x_S|X_K) \right]=|\Xc|^{-ts},
\end{align*}
and the expected successful guessing probability after observing $Y$ is 
\begin{align*}
{\rm P_{\rm s}}(X_K,Y)={\mathbb E}_{Y,X_K} \left[ \max_{x_S} P_{X_S|Y,X_K}(x_S|Y,X_K) \right].
\end{align*}
The leakage $\leak$ is defined as the logarithm of the ratio between the expected probabilities of the adversary successfully guessing $x_Q$ {after and before} observing $Y$ \cite{smith2009foundations,braun2009quantitative}. That is, 
\begin{align}
\leak&\doteq \log_q \frac{{\rm P_{\rm s}}(X_K,Y)}{{\rm P_{\rm s}}(X_K)}  \label{eq:def:leakage:rate:any:t:f:mid}  \\
&=\log_q \sum_{x_K,y} \max_{x_S} P_{Y,X_K|X_S}(y,x_K|x_S).  \label{eq:def:leakage:rate:any:t:f}
\end{align}
%
%
The leakage rate of the code is $\Lc=t^{-1} \leak$ and the optimal leakage rate can then be defined as 
\begin{align}
\Lc^*&\doteq \lim_{t\to \infty} \inf_{\text{$(t,M,f,\gv)$ codes}} \Lc.  \label{eq:def:leakage:rate} 
\end{align}
%
%

\begin{remark}
The idea of measuring leakage as the ratio of the adversary's successful guessing probabilities has been introduced and explored in various contexts \cite{smith2009foundations,braun2009quantitative,issa2019operational}. 
The leakage $\leak$ is equal to the maximal leakage \cite{issa2019operational} from $X_{S}$ to $Y$ given side information $X_K$, which is also equal to the maximum min-entropy leakage \cite{braun2009quantitative} from $X_{S}$ to $Y$ given $X_K$. 
\end{remark}

\section{A Deterministic Linear Index Code}  \label{sec:achievability}

%
%

In the following, we construct a deterministic linear index code based on the minrank method and fitting matrices, which were developed for the original index coding problem without adversary \cite{bar2011index}. 
We then show that the proposed scheme achieves the optimal leakage rate over all valid deterministic scalar linear index codes. 
Throughout the section, we set $t=1$ as we are considering only scalar linear codes.

Unless otherwise stated, we use bold-faced capital letters to denote matrices and vectors, e.g., 
$\Xv_{[n]}=[X_1 \enskip \ldots \enskip X_n]^T$. 
Let $r(\cdot)$ denote the rank of a matrix over the Galois field $\Fb_q$.

Note that for any receiver $i$ requiring more than one messages (i.e. $|W_i|\ge 2$), we can transform the problem into a new equivalent problem by removing the receiver $i$ and adding $|W_i|$ new receivers, where every new receiver has the same side information set $A_i$ and each receiver wants a unique message in set $W_i$. 
Therefore, we can, without loss of generality, always assume $W_i=\{w_i\}$ being a singleton set for every receiver $i\in [m]$. 

For any given problem, a size $m\times n$ fitting matrix $\Mv$ of Galois field $\Fb_q$ is a matrix such that for any receiver $i\in [m]$, 
\begin{align}
\Mv_{ij}&=1, \quad \text{for $j=w_i$,}  \\
\Mv_{ij}&=0, \quad \text{for any message $j\in [n]\setminus (W_i\cup A_i)$}.
\end{align}
As $\Mv_{ij}$ can be any element in $\Fb_q$ if $j\in A_i$, there can be multiple fitting matrices for a given problem. 
For example, for the problem with $n=5$ messages and $m=3$ receivers, where $W_i=\{i\},\forall i\in [3]$, $A_1=\{3,4\}$, $A_2=\{1,4,5\}$, $A_3=\{2,5\}$, any matrix of the form below is a fitting matrix, 
\begin{align}  \label{eq:fitting:matrix:group}
\begin{bmatrix}
1 & 0 & ? & ? & 0\\
? & 1 & 0 & ? & ?\\
0 & ? & 1 & 0 & ?
\end{bmatrix}
\end{align}
where ``?" means that the entry can be any element in $\Fb_q$. 

If the server generates codeword $\Yv$ by multiplying a fitting matrix $\Mv$ by the message vector $\Xv_{[n]}=[X_1 \enskip X_2 \enskip \ldots \enskip X_n]^T$, every receiver $i\in [m]$ can recover its wanted message because the $i$-th element of $\Yv$ is a linear combination of only $X_{w_i}$ and some of its side information. 
%
Moreover, note that any row of $\Mv$ can be generated by $r(\Mv)$ independent rows of $\Mv$. Thus, 
to satisfy the decoding requirements at the receivers, the server needs only to transmit the linear combination of the messages with coefficients from $r(\Mv)$ independent rows of $\Mv$. 
In this way, for a given problem, the minimum rank over all the fitting matrices, namely, the \emph{minrank} value, establishes an upper bound on its broadcast rate, which has been proved to be optimal over all deterministic scalar linear codes \cite{bar2011index}. 

For analysis of information leakage based on the fitting matrix framework, 
we can split $\Mv$ into three submatrices formed by different groups of columns in $\Mv$ according to sets $K$, $S$, and $U$. For brevity, we simply write 
\begin{align}
\Mv=[\Kv \quad \Sv \quad \Uv].
\end{align}

The following theorem characterizes the minimal leakage rate among all codes based on the fitting matrix framework.

\begin{theorem}  \label{thm:leakage:minrank:achievability}
For any index coding problem, there exists a deterministic scalar linear index code that yields the following leakage rate, 
\begin{align}
\Lc=\min_{\Mv} (r([\Sv \quad \Uv])-r(\Uv)).  \label{eq:matrix:leakage}
\end{align}
Furthermore, this result is leakage-wise rate optimal for all deterministic scalar linear codes. 
\end{theorem}

\begin{IEEEproof}
We first show the achievability. 
Consider any fitting matrix $\Mv=[\Kv \quad \Sv \quad \Uv]$ and any encoding matrix $\Ev$ formed by a set of row vectors of $\Mv$ such that the row space of $\Mv$ is the same as that of $\Ev$ and thus by receiving $\Yv=\Ev \Xv$ every receiver can decode its wanted message. 
As $t=1$, we have 
\begin{align*}
\Lc&=\log_q \sum_{x_K,y} \max_{x_S} P_{Y,X_K|X_S}(y,x_K|x_S)  \\
&=\log_q \sum_{x_K,y} \max_{x_S} \sum_{x_U} P_{Y,X_K,X_U|X_S}(y,x_K,x_U|x_S)  \\
&\stackrel{(a)}{=}\log_q \sum_{x_K,y} \max_{x_S} \sum_{x_U} P_{Y|X_{[n]}}(y|x_{[n]}) \cdot P_{X_{K\cup U}}(x_{K\cup U})  \\
&\stackrel{(b)}{=} \log_q \big( q^{-k-u} \sum_{x_K} \sum_{y} \max_{x_S} \sum_{x_U} {\mathbbm{1}}( \Yv = \Ev \Xv) \big)  \\
&\stackrel{(c)}{=} \log_q \big( q^{-k-u} \sum_{x_K} \sum_{y} q^{u-r(\Uv)} \big)  \\
&\stackrel{(d)}{=}r([\Sv \quad \Uv])-r(\Uv),
\end{align*}
where (a) is due to the messages being independent, 
(b) is due to the code being deterministic, where ${\mathbbm{1}}(\cdot)$ is the indicator function, 
(c) follows from the fact that given any fixed $y$ and $x_{K\cup S}$, there are $q^{u-r(\Uv)}$ possible $x_U$ values that satisfy $\Yv=\Ev\Xv$, 
and (d) follows from the fact that given any fixed $x_K$, there are $q^{r([\Sv \quad \Uv])}$ possible $y$ values. 
Therefore, by minimizing over all fitting matrices, the leakage rate in \eqref{eq:matrix:leakage} can be achieved. 

Now we prove the converse part of the theorem. 
Suppose a $\ell\times n$ matrix $\Evt$ of Galois field $\Fb_q$ 
is the encoding matrix of an arbitrary valid deterministic scalar linear index code. 
Note that $\Evt$ need not be a fitting matrix. 
%
We split $\Evt$ into three submatrices formed by different groups of columns according to sets $K$, $S$, and $U$ as
\begin{align}
\Evt=[\Kvt \quad \Svt \quad \Uvt].
\end{align}
Following a similar argument as in the achievability proof of the theorem, we can show that the leakage rate caused by the codeword $\Yv=\Evt \Xv$ is 
\begin{align}
\Lc_{\Evt}=r([\Svt \quad \Uvt])-r(\Uvt).  \label{eq:midmidmid}
\end{align}
It remains to show that $L_{\Evt}$ is lower bounded by \eqref{eq:matrix:leakage}. 

According to the proof of \cite[Theorem 1]{bar2011index}, there exists some fitting matrix $\Mv=[\Kv \quad \Sv \quad \Uv]$ of the problem such that the row vectors of $\Mv$ lie in the row space of $\Evt$. 
%
In other words, there exists some $n\times \ell$ matrix $\Bv$ such that $\Bv \Evt = \Mv$, or, if we only consider the submatrices according to sets $S$ and $U$, $\Bv [\Svt \quad \Uvt]=[\Sv \quad \Uv]$. 
Also, there exists some matrix $\Dv$ such that $\Uvt=[\Svt \quad \Uvt] \Dv$, $\Uv=[\Sv \quad \Uv] \Dv=\Bv [\Svt \quad \Uvt] \Dv$. 
Hence, 
%
\begin{align}
r([\Sv \quad \Uv])+r(\Uvt)
&=r(\Bv [\Svt \quad \Uvt])+r([\Svt \quad \Uvt] \Dv)  \nonumber  \\
&\stackrel{(a)}{\le} r( [\Svt \quad \Uvt])+r(\Bv  [\Svt \quad \Uvt] \Dv)  \nonumber  \\
&=r( [\Svt \quad \Uvt])+r(\Uv),  \nonumber
\end{align}
where 
(a) is due to Frobenius inequality \cite{baksalary2013k}. By reorganizing the above result, we have 
$\Lc_{\Evt}=r([\Svt \quad \Uvt])-r(\Uvt)\ge r([\Sv \quad \Uv])-r(\Uv),$ 
which, together with the fact that $\Mv$ is a fitting matrix, completes the proof. 
\end{IEEEproof}

\begin{remark}
Following similar arguments as in the proof of Theorem \ref{thm:leakage:minrank:achievability}, we can show that when the mutual information\footnote{Note that mutual information between sensitive variables and codeword has been commonly used as a leakage metric in the literature \cite{shannon1949communication,yamamoto1994coding,schieler2014rate,shkel2020compression,guo2020information}.} $I(X_S;Y|X_K)$ is used as the leakage metric, $\min_{\Mv}(r([\Sv \quad \Uv])-r(\Uv))$ still characterizes the minimal leakage rate over all deterministic scalar linear index codes. 
It seems that the rank $r(\Uv)$ can be viewed as a rough measure of the level of protection against the adversary provided by the non-sensitive messages $X_U$.
\end{remark}

\begin{remark}
Computing \eqref{eq:matrix:leakage} for index coding instances with a large number of messages and receivers can be computationally challenging. 
When there is no non-sensitive messages, computing \eqref{eq:matrix:leakage} reduces to computing the {minrank} value for the index coding instance over the finite field $\Fb_q$, which has been shown to be NP-complete \cite{el2007minimum}. 
\end{remark}

The example below shows the efficacy of the scheme. 

\begin{example}  \label{exm:linear:1}
Consider the problem $\Gamma$ with $n=5$ binary messages and $m=5$ receivers with $W_i=\{i\},i\in [m]$ and $$ A_1=\{4,5\},A_2=\{1\},A_3=\{2\},A_4=\{3\},A_5=\{4\}. $$
Assume for the adversary, $K=\{5\}, S=\{1,3\}, U=\{2,4\}$.
The fitting matrix 
\begin{align}  \label{eq:exm:linear:1}
\Mv=
\begin{bmatrix}
1 & 0 & 0 & 1 & 0\\
1 & 1 & 0 & 0 & 0\\
0 & 1 & 1 & 0 & 0\\
0 & 0 & 1 & 1 & 0\\
0 & 0 & 0 & 0 & 1
\end{bmatrix}
\end{align}
achieves the broadcast rate of $\beta=r(\Mv)=4$. It also gives 
\begin{align*}
\Lc=r([\Sv \quad \Uv])-r(\Uv)=3-2=1,
\end{align*}
which is indeed the optimal leakage rate as we will show in Example \ref{exm:converse:tight} in the next section. 
The linear code given by \eqref{eq:exm:linear:1} is optimal in both leakage and compression senses. 
\end{example}

However, as we show by the simple two-receiver example below, it is not always possible to simultaneously achieve the optimal compression and leakage rates. 

\begin{example}  \label{exm:linear:2}
Consider the problem $\Gamma$ with $n=4$ binary messages and $m=2$ receivers, where 
$$ W_1=\{1\}, W_2=\{2\}, A_1=\{2,3\}, A_2=\{1,4\}.$$
Assume for the adversary, $K=\emptyset, S=\{1,2\}, U=\{3,4\}$. 
The fitting matrix $\Mv=
\begin{bmatrix}
1 & 1 & 0 & 0\\
1 & 1 & 0 & 0\\
\end{bmatrix}$
achieves the broadcast rate of $\beta=r(\Mv)=1$ while resulting in a leakage rate of $\Lc=r(\Mv)-r(\Uv)=1-0=1$. 
The following fitting matrix 
$\Mvt=
\begin{bmatrix}
1 & 0 & 1 & 0\\
0 & 1 & 0 & 1\\
\end{bmatrix}$
gives $\Lc=r(\Mvt)-r(\Uvt)=2-2=0$, 
indicating that zero leakage (i.e., perfect secrecy) can be achieved for the problem. 
However, $\Mvt$ leads to a suboptimal compression rate of $r(\Mvt)=2$. 
In fact, in the following we use Shannon-type inequalities \cite[Chapter 14]{yeung2008information} to show that the compression rate of any index code that attains zero leakage is at least $2$. 
Recall that we are considering binary uniform messages that are independent to each other. Thus, $t=1$ and $H(X_{J})=|J|,\forall J\subseteq [4]$. 
Consider any $(t,M,f,\gv)$ index code that achieves zero leakage, i.e., $L=I(X_{\{1,2\}};Y)=0$. 
We have $2=H(X_{\{1,2\}})=H(X_{\{1,2\}}|Y)=H(X_1|Y)+H(X_2|Y,X_1)=H(X_2|Y)+H(X_1|Y,X_2)$, 
which implies that $H(X_2|Y,X_1)=H(X_1|Y,X_2)=1$. 
Then, we have 
\begin{align*}
&H(X_3|Y,X_{\{1,2\}})  \\
&=H(X_3|Y,X_2)-H(X_1|Y,X_2)+H(X_1|Y,X_{\{2,3\}})  \\
&\stackrel{(a)}{=}H(X_3|Y,X_2)-1\le H(X_3)-1=0,
\end{align*}
where (a) follows since that $H(X_1|Y,X_2)=1$ and $H(X_1|Y,X_{\{2,3\}})=0$ due to the decoding requirement that receiver $1$ must be able to recover $X_1$ from $Y$ and $X_{\{2,3\}}$. 
As entropy is always non-negative, we have $H(X_3|Y,X_{\{1,2\}})=0$. 
Similarly, we can show that $H(X_4|Y,X_{\{1,2\}})=0$ and hence $H(X_{\{3,4\}}|Y,X_{\{1,2\}})=0$. 
Thus, we have 
\begin{align*}
t^{-1}\log_q M&\ge H(Y)  \\
&\ge I(Y;X_{\{3,4\}}|X_{\{1,2\}})  \\
&=H(X_{\{3,4\}}|X_{\{1,2\}})-H(X_{\{3,4\}}|Y,X_{\{1,2\}})  \\
&=H(X_{\{3,4\}})-H(X_{\{3,4\}}|Y,X_{\{1,2\}})=2.
\end{align*}
That is, the compression rate of any index coding attaining zero leakage must be at least $2$. 
Therefore, $\beta=1$ and $\Lc^*=0$ can never be simultaneously achieved for this problem. 
\end{example}

\section{A General Lower Bound on $\Lc^*$}  \label{sec:converse}
We derive the following lower bound on the leakage rate that holds generally over any valid index codes.

\begin{theorem}  \label{thm:pqs:converse}
For any index coding problem $\Gamma$, we have 
\begin{align}
\Lc^*(\Gamma)
\ge \beta({\tilde \Gamma}(S\cup U))-u,
\end{align}
where ${\tilde \Gamma}$ denotes the index coding problem constructed by adding an extra receiver, indexed by $m+1$, to the original problem, which knows side information $A_{m+1}=K\cup S$ and wants messages $W_{m+1}=U$. 
\end{theorem}



\begin{IEEEproof}
Consider any valid $(t, M, f, \gv)$ index code. 
For any $(x_K,y)$ value and any $J\subseteq S\cup U$, 
let $$\Xc_{J}(x_K,y)=\{ x_{J}\in \Xc_{J}:P_{Y,X_{K\cup J}}(y,x_{K\cup J})>0 \}$$ denote the set of $X_{J}$ values jointly possible with $(x_K,y)$. 

Now consider the problem ${\tilde \Gamma}$, which is constructed from $\Gamma$ by adding an extra receiver $m+1$ with $A_{m+1}=K\cup S$ and $W_{m+1}=U$.  
Notice that by adding a receiver we are essentially adding more edges into the confusion graph (i.e., $E(\Gamma_t)\subseteq E({\tilde \Gamma_t})$). 
We first show the following inequality: 
\begin{align}
\max_{x_K,y}|\Xc_S(x_K,y)|\le \alpha({\tilde \Gamma}_t(S\cup U)),  \label{eq:pqs:converse:key}
\end{align} 
where $\alpha(\cdot)$ denotes the independence number of a graph. 

We prove the inequality via contradiction. 
Assume there exists some $(x_K,y)$ value such that $|\Xc_S(x_K,y)|>\alpha({\tilde \Gamma}_t(S\cup U))$. 
Consider a subset $\Ic$ of the vertex set $\Xc_{S\cup U}(x_K,y)$ in the subgraph $\Gamma_t(S\cup U)$ such that for every $x_S\in \Xc_S(x_K,y)$, there is exactly one vertex in $v_{S\cup U}\in \Ic$ such that $v_S=x_S$. 
Thus $|\Ic|=|\Xc_S(x_K,y)|$. 
For a visualization of the construction of $\Ic$, see the schematic graph in Figure \ref{fig:pqs:converse}. 


\begin{figure}[ht]
\begin{center}
\includegraphics[scale=0.225]{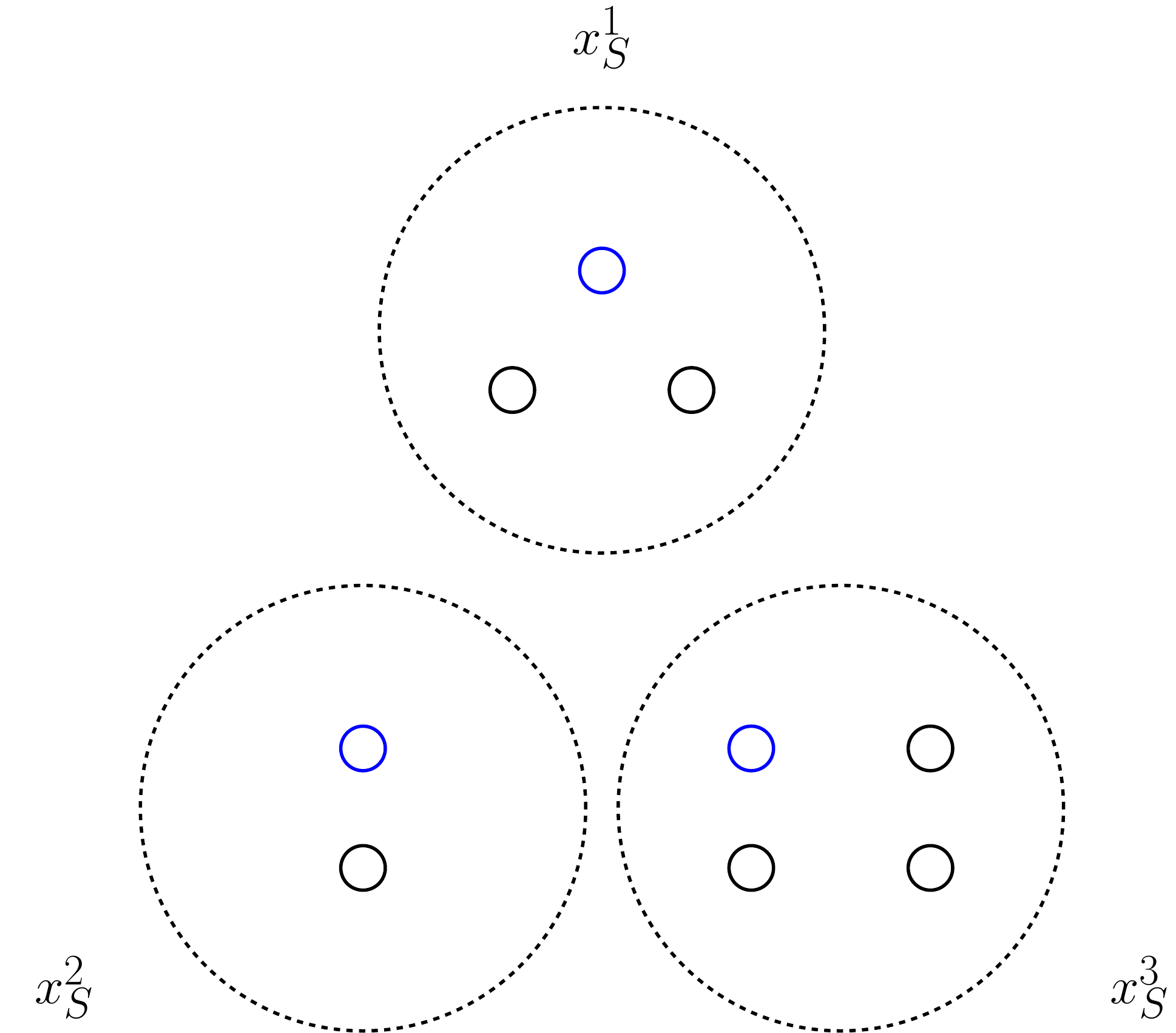}
\caption{
Vertices in $\Xc_{S\cup U}(x_K,y)$ are denoted by nodes in the figure. 
Note that the figure is for illustrative purpose only; there is no limit on the number of vertices in $\Xc_{S\cup U}(x_K,y)$. 
We partition the nodes into subgroups according to their $x_S$ values, and each subgroup is denoted by a dashed circle with its corresponding $x_S$ value marked beside it. 
To construct $\Ic$, one needs only to arbitrarily pick one node from each dashed circle. 
For example, $\Ic$ can be constructed by the blue nodes, one from each subgroup. 
The number of distinct $x_S$ values in $\Xc_{S\cup U}(x_K,y)$, $|\Xc_S(x_K,y)|$, is equal to the number of dashed circles in the graph, which is then equal to $|\Ic|$. 
}
\label{fig:pqs:converse}
\end{center}
\end{figure}


By the definition of the confusion graph, $\Xc_{S\cup U}(x_K,y)$ is an independent set in the induced subgraph $\Gamma_t(S\cup U)$. 
Since $\Ic\subseteq \Xc_{S\cup U}(x_K,y)$, $\Ic$ is also an independent set in $\Gamma_t(S\cup U)$. 
Thus, any two vertices in $\Ic$ are not confusable at any receiver $i\in S\cup U$ of the subproblem $\Gamma(S\cup U)$. 
Note that the extra receiver $m+1$ knows $X_S$ as side information. 
Since any two vertices in $\Ic$ have different $x_S$ values by construction, they are not confusable at the extra receiver $m+1$ either. 
In conclusion, any two vertices in $\Ic$ are not confusable at any receivers in the subproblem ${\tilde \Gamma}(S\cup U)$. 
In other words, any two vertices in $\Ic$ are not adjacent in ${\tilde \Gamma}_t(S\cup U)$, and hence $\Ic$ is also an independent set of ${\tilde \Gamma}_t(S\cup U)$. 
Therefore, we must have $|\Ic|\le \alpha({\tilde \Gamma}_t(S\cup U))$, which contradicts with the assumption that $|\Xc_S(x_K,y)|=|\Ic|>\alpha({\tilde \Gamma}_t(S\cup U))$.

Having proved \eqref{eq:pqs:converse:key}, we have 
\begin{align}
&\sum_{x_K,y} \max_{x_S} P_{Y,X_{K,S}}(y,x_{K,S})  \nonumber  \\
&\stackrel{(a)}{\ge} \sum_{x_K,y} \frac{1}{|\Xc_{S}(x_K,y)|} \sum_{x_S\in \Xc_{S}(x_K,y)} P_{Y,X_{K,S}}(y,x_{K,S})  \nonumber  \\
&\stackrel{(b)}{\ge} \frac{1}{\alpha({\tilde \Gamma}_t(S\cup U))} \sum_{x_K,y} \sum_{x_S\in \Xc_{S}(x_K,y)} P_{Y,X_{K,S}}(y,x_{K,S})  \nonumber  \\
&=\frac{1}{\alpha({\tilde \Gamma}_t(S\cup U))},  \label{eq:pqs:converse:mid}
\end{align}
where (a) follows since the maximum is larger than the average, and (b) is due to \eqref{eq:pqs:converse:key}. 
Finally, we have 
\begin{align}
&\Lc^*(\Gamma)  \nonumber  \\
&=\lim_{t\to \infty} \frac{1}{t} \min_{\text{$(t,M,f,\gv)$ codes}} \log_q \sum_{x_K,y} \max_{x_S} P_{Y,X_{K}|X_S}(y,x_{K}|x_S)  \nonumber  \\
&\stackrel{(c)}{\ge} \lim_{t\to \infty} \frac{1}{t} \log_q \frac{|\Xc|^{ts}}{\alpha({\tilde \Gamma}_t(S\cup U))}  \nonumber  \\
&\stackrel{(d)}{=} \lim_{t\to \infty} \frac{1}{t} \log_q \frac{|V({\tilde \Gamma}_t(S\cup U))|\cdot |\Xc|^{-tu}}{\alpha({\tilde \Gamma}_t(S\cup U))}  \nonumber  \\
&\stackrel{(e)}{=} \lim_{t\to \infty} \frac{1}{t} \log_q \chi_{\rm f}({\tilde \Gamma}_t(S\cup U)) - u  \nonumber  \\
&\stackrel{(f)}{=} \beta({\tilde \Gamma}(S\cup U))-u,  \nonumber
\end{align}
where (c) follows from \eqref{eq:pqs:converse:mid} and the fact that all $x_S$ are equally likely, (d) follows from $|V({\tilde \Gamma}_t(S\cup U))|=|\Xc|^{t(s+u)}$, (e) follows since confusion graphs are vertex-transitive \cite{scheinerman2011fractional}, and (f) follows from \eqref{eq:compression:zero:error}. 
\end{IEEEproof}

\begin{remark}
While in the current work ${\tilde \Gamma}$ only appears as a part of the computable expression of the converse result, the deeper relationship between $\Gamma$ and ${\tilde \Gamma}$ in both compression and secrecy senses is worth investigating. 
\end{remark}

\begin{remark}
It can be shown using the generalized maximal acyclic induced subgraph (MAIS) bound \cite{bar2011index} that the lower bound in Theorem \ref{thm:pqs:converse} is always non-negative. 
In particular, when there are no non-sensitive messages in the system (i.e., $U=\emptyset$), the extra receiver in ${\tilde \Gamma}(S\cup U)$ becomes trivial as it knows all the messages in the system and wants nothing. In such case, the lower bound in Theorem \ref{thm:pqs:converse} reduces to that in Corollary 1 in our previous work \cite{itw2021leakage}, and is always tight. 
\end{remark}

Theorem \ref{thm:pqs:converse} can sometimes yield tight lower bounds on $\Lc^*$. 
\begin{example}  \label{exm:converse:tight}
Consider the problem $\Gamma$ in Example \ref{exm:linear:1}. Theorem \ref{thm:pqs:converse} gives 
$\Lc^*(\Gamma)\ge \beta({\tilde \Gamma}(\{1,2,3,4\}))-|\{2,4\}|=3-2=1,$
where $\beta({\tilde \Gamma}(\{1,2,3,4\}))=3$ can be proved using the MAIS bound \cite{bar2011index} and the cycle covering scheme \cite{chaudhry2011complementary,neely2013dynamic}. 
The lower bound on $\Lc^*(\Gamma)$ above matches the achievability result in Example \ref{exm:linear:1} and thus establishes $\Lc^*(\Gamma)$. 
\end{example}


\begin{remark}
It can be shown utilizing existing results on broadcast rates \cite[Theorem 1.1]{lubetzky2009nonlinear} that no constant gap exists between the upper and lower bounds in Theorems \ref{thm:leakage:minrank:achievability} and \ref{thm:pqs:converse}. 
\end{remark}

\newpage

\bibliographystyle{IEEEtran}
\bibliography{references}

\end{document}